\documentclass[prd,superscriptaddress,onecolumn,showkeys]{revtex4}
\usepackage{textcomp}
\usepackage{eurosym}
\usepackage{amsfonts}
\usepackage{array}
\usepackage{amsthm}
\usepackage{bm}
\usepackage{palatino}
\usepackage[colorinlistoftodos]{todonotes}
\usepackage{mathpazo}
\usepackage{supertabular}
\usepackage{subfig}
\usepackage{amssymb}
\usepackage{eurosym}
\usepackage{amsmath}
\usepackage{epsfig}
\usepackage{graphics}
\usepackage{changes}
\usepackage{color}
\usepackage{graphicx}
\usepackage[colorlinks=true,
            linkcolor=blue,
           urlcolor=black,
           citecolor=black]{hyperref}

\def\be{\begin{equation}}
\def\ee{\end{equation}}
\def\beq{\begin{eqnarray}}
\def\eeq{\end{eqnarray}}

\def\bes{\begin{eqnarray}}
\def\ees{\end{eqnarray}}

\begin{document}
\title{Study of tunneling radiation and thermal fluctuations of a gauge super gravity like black hole}

\author{Riasat Ali}
\email{riasatyasin@gmail.com}
\affiliation{Department of Mathematics, GC
University Faisalabad Layyah Campus, Layyah-31200, Pakistan}

\author{Muhammad Asgher}
\email{m.asgher145@gmail.com}
\affiliation{Department of Mathematics, The Islamia
University of Bahawalpur, Bahawalpur-63100, Pakistan}

\author{P.K. Sahoo}
\email{pksahoo@hyderabad.bits-pilani.ac.in}
\affiliation{Department of Mathematics, Birla Institute of Technology and Science-Pilani, Hyderabad Campus, Hyderabad-500078, India}

\date{\today}

\begin{abstract}
 The Newman-Janis technique and the semi-classical Hamilton-Jacobi approach are two distinct phenomena that we apply in this paper to examine the Hawking temperature $(T_{H})$ for $4$-dimensional gauge super gravity like black hole with rotation parameters. First, using the Newman-Janis algorithmic approach, we compute the gauge super gravity like black holes solution. We use surface gravity to derive the $T_H$ for a gauge super gravity like black hole. In order to do this, we use the Lagrangian field equation in the background of generalized uncertainty principle, semi-classical Hamilton-Jacobi approach and the WKB approximation technique. Furthermore, we investigate the stability of considered geometry with the help of corrected entropy and well known thermodynamic quantities. It is concluded that the gauge super gravity like black holes is stable under first-order corrections.
\end{abstract}


\date{\today}

\maketitle

\section{Introduction}

The Janis-Newman method is presented in its most general form of BH geometry. With the help of this extension, configurations with all bosonic fields having spin of one can be created complex and real scalar fields, gauge fields and metric fields. The Newman-Penrose tetrad formalism has been used in the original prescription, which seems to be exceedingly time-consuming because it calls for inverting the metric, looking for a basis of null tetrad where the transformation may be used, and then inverting into metric \cite{1,2,2a,3,4, 4a}.
One of these solution generating methods has the Janis-Newman phenomenon, which, in its original form, has derived rotational metrics from static ones.
It was discovered by Janis and Newman as an alternate Kerr metric formulation \cite{5}, and soon after, it was utilised once more to find the Kerr-Newman metric \cite{6}. The Kerr-Newman metric is the only rotational solution the Janis-Newman algorithm has produced for non-fluid configurations (excluding radiated and internal solutions) over the course of its long history, and only few known cases have been created \cite{7,8,9}. In general, applying the Janis-Newman phenomenon to interior and emitting systems \cite{10,11,12,13,14} entails deriving a configuration that does not immediately solve the motion equations and interpreting the fluid (whose features can be analysed); however, these types of applications are not the focus of this review.

The generalized uncertainty principle (GUP) is the most appropriate technique to investigate this minimum length \cite{GUP1}. A definition of the modified commutation relationship is
\begin{equation}
 [ O_{u}, p_{v}]=i\hbar\left[1+\alpha(p^{2})\right]\delta_{uv},
\end{equation}
with $p_{v}$ and $O_{u}$ are generalized momentum and position operators, respectively. Moreover, a connection to GUP can be made in the following manner
\begin{equation}
\Delta O\Delta p\geq \frac{\hbar}{2}\left[1+\alpha(\Delta p^{2})\right].
\end{equation}
The Planck mass is denoted by $M_{p}$, and the correction parameter $\alpha$ can be represented in terms of a dimensionless parameter as $\alpha=\frac{\alpha_{0}}{M_{p^{2}}}$. The quantum effects have a significant impact close to a BH's horizon and the GUP relationship is highly helpful in understanding the BH physics.

Hawking investigated the theory that black holes (BHs) emit radiation around the horizon as a result of the effects of the associated quantum vacuum fluctuation \cite{15}.
The quantum tunneling method can be used to define Hawking radiation near to the BHs horizon.
It was discovered that the black body radiation was the Hawking radiation.
This scenario suggests that the BH totally \cite{16} disappears.
Different inferences based on relativity quantum mechanics and quantum theory in curved space-time \cite{17} demonstrate the accuracy of the Hawking formula.
The semi-classical tunnelling strategy \cite{18,19,20,21,22,23,24,25,26,27,28,28a} is a useful model for investigating the Hawking radiation. One can calculate the corrected $T_{H}$ using this method by accounting for the different space-time.
All sorts of particles, including fermions, bosons, and scalar particles, are radiated by the generalised BH.
For these kinds of particles in the geometry of various BHs metrics, the standard $T_{H}$ as well as the corrected $T_{H}$ have been examined \cite{29,30,31}.
The fact that the corrected $T_{H}$ rate is higher than the standard $T_{H}$ indicates that the evaporation process is accelerated by the various background geometries. There are mentions to a number of commutation relations generalisations \cite{32,33,34}. These alterations are widely used in order to discover more about the quantum features of gravity. Black holes can be used to examine the effects of quantum gravity. The introduction of gravity influences into physics of BH through the use of GUP leads to several interesting outcomes and discoveries \cite{30}.
Additionally, the analysis included the remnant mass and modifications to the entropy and temperature. The effects of GUP to spin-$0$ and spin-$1$ particles in the context of warped DGP gravity BH have been studied \cite{35} by \"{O}vg\"{u}n and Jusufi.

Thermal fluctuations are caused by statistical perturbations in compact objects. Because Hawking radiations are emitted from BHs, they reduce the size of the BH and, as a result, raise the temperature of the considered BH. Faizal and Khalil \cite{z3} looked on the thermodynamics and thermal fluctuations of charged AdS BHs. This type of BH shows the stability under the first-order corrections. Pourhassan and Faizal \cite{z4} investigated the thermodynamics of spinning Kerr-AdS BH and they studied the corrected entropy and analyzed that the corrections are extremely effective for the small BHs. Additionally, utilizing heat capacity and the Hessian matrix stability test, the non-minimal regular BHs of thermodynamics are studied \cite{z5}. They came to the conclusion that the these BHs are locally and globally stable for increasing levels of the correction parameter. Zhang and Pradhan \cite{z6} investigated that how corrected entropy affected the Kerr-Newman-AdS and RN-AdS BHs, they found that there was a quadratic phase transition occurring because of the thermal fluctuations.
Moreover, the thermodynamics  of various types of BHs under the influence of correction parameter is analyzed, they checked the relation of other thermodynamics quantities with the corrected entropy and for stability reasons \cite{z13,z14a,z14b,z14c,z15,z16,z17,z18, RB1, RB2}.

In this article, we examine the Newman-Janis method that generates rotating metrics by performing a complex transformation. In section  \textbf{II}, we develop the rotating metric for a spherically symmetric metric that will be used to analyse the $T_H$. In section.  \textbf{III} generates both corrected tunnelling solutions and corrected $T_H$. Graphics are used in Sec. \textbf{IV} to describe the solution's physical behaviour, viability, and stability. In section \textbf{V}, we investigate the behaviour of thermodynamic quantities under the effect of thermal fluctuations. We finalise our results in section \textbf{VI}.

\section{Gauged Super Gravity Like Black Hole with Rotation Parameter}

The theory of representation for the extended algebra of supersymmetry studied by Wess and Bagger \cite{S1} in
chapter 2. Gibbons \cite{S2} studied the Reissner-Nordstrom BH as a supersymmetric solition. The supergravity models of BH solutions have been studied \cite{S3} and also discussed the general BHs solution in supersymmetric theories and general relativity.
Black holes are significant objects in any gravity theory for the insight  BHs implies
 into quantum gravity. For this reason, obtaining all potential BH solutions is an important stage in any theory. Since it is thought that the majority of astrophysical BHs are revolving, rotating BHs are the most pertinent sub-cases for astrophysics. These systems might also include revolving star external metrics.
This algorithm implies an approach to generating an axi-symmetric solution from a seed solution that is spherically symmetric by first complexifying the radial and null time coordinates, then performing a complex coordinate transformation.
To write the outcome in Boyer-Lindquist(BL) coordinates, one frequently changes the spherical coordinates. The gauge super gravity like BH is given as \cite{24, 36},

\begin{equation}
ds^{2}=-F(r)dt^2+\frac{1}{F(r)}dr^2+r^2G(r)( d\theta^2+ sin^2 \theta^2d\phi^2),\label{d1}
\end{equation}
where $F(r)=\frac{f}{(H_{1}H_{2}H_{3}H_{4})^{\frac{1}{2}}}$ and $G(r)=(H_{1}H_{2}H_{3}H_{4})^{\frac{1}{2}}$,
with $f=k-\frac{\mu}{r^2}+g^2r^2H_{1}H_{2}H_{3}H_{4},~H_{i}=\frac{q_{i}}{r^2}+1$, for $(i=1,2,3,4).$
 The $g = 1/L$ and $L$ is related to the cosmological constant $\Lambda =-3g^2=-3/L^2$ and the $\mu$ represent
the non-extremality parameter \cite{37} for radius $k=1$ and $k=0,$ then represents the metrics on $S_{2}$,
and $R_{2}$ respectively. The four electric potentials $A_{\mu}^{i}$ are defined as;
 $A_{0}^{i}=\frac{\tilde{q_{i}}}{r^2+q_{i}}$ $(i=1,2,3,4),$ where $q_{i}$ and $\tilde{q_{i}}$ represent
 charges and physical charges of a BH. Through the transformation of
$(t, r, \theta, \phi)$ to $(u, r, \theta, \phi)$ coordinates
\begin{eqnarray}
du=dt-\frac{dr}{F(r)},\label{A}
\end{eqnarray}

After using the above transformation the metric (\ref{d1}) can be expressed as follows
\begin{equation}
ds^{2}=-F(r)du^2-2dudr+r^2 G(r)( d\theta^2+ sin^2d\phi^2).\label{A5}
\end{equation}  
The non-zero component of the above metric is
\begin{eqnarray}
g_{uu}&=&-F(r),~g_{ur}=g_{ru}=-1,
~~g_{\theta\theta}=G(r)r^2,\nonumber\\
g_{\phi\phi}&=& G(r)r^{2}\sin^2\theta.
\end{eqnarray}
Now, we take the inverse metric coefficient
\begin{eqnarray}
g^{rr}&=&-F(r),~~
g^{\theta\theta}=\frac{1}{G(r)r^2},~~g^{\phi\phi}=\frac{1}{G(r)r^2sin^2\theta},\nonumber\\
g^{ur}&=&g^{ru}=-1.
\end{eqnarray}
In the null tetrad framework the metric can be written as
\begin{eqnarray}
g^{\mu\nu}=-l^\nu n^\mu-l^\mu n^\nu+m^\mu \bar{m}^{\nu}+m^\nu \bar{m}^{\mu}.\label{AA6}
\end{eqnarray}
Here the corresponding components are
\begin{eqnarray}
l^{\mu}&=&\delta_{r}^{\mu},~~~n^{\nu}=\delta_{u}^{\mu}-\frac{1}{2} F \delta_{r}^{\mu},\nonumber\\
m^{\mu}&=&\frac{1}{\sqrt{2}r G(r)}( \delta_{\theta}^{\mu}+\frac{i}{sin\theta}\delta_{\phi}^{\mu}),\nonumber\\
\bar{m}^{\mu}&=&\frac{1}{\sqrt{2}r G(r)} (\delta_{\theta}^{\mu}-\frac{i}{sin\theta}\delta_{\phi}^{\mu}).\nonumber
\end{eqnarray}
For any point in the BH spacetime the null vectors of the null tetrad satisfy the relations of
$l_{\mu}l^{\mu}=n_{\mu}n^{\mu}=m_{\mu}m^{\mu}=l_{\mu}m^{\mu}=m_{\mu}m^{\mu}=0$
and $l_{\mu}n^{\nu}=-m_{\mu}\bar{m}^{\mu}=1$ and  $(u, r)$ place coordinate transformation are
$u\rightarrow u-iacos\theta$, $r\rightarrow r+iacos\theta$, then we perform the transformation
$F(r)\rightarrow \tilde{F}(r, a, \theta)$ and $\sigma^2=r^2+a^2 cos^2\theta$.
In the $(u, r)$ space the vectors become
\begin{eqnarray}
l^{\mu}&=&\delta_{r}^{\mu},~~~n^{\nu}=\delta_{u}^{\mu}-\frac{1}{2}
\tilde{F} \delta_{r}^{\mu},\nonumber\\
m^{\mu}&=&\frac{1}{\sqrt{2}r G(r)}(\delta_{\theta}^{\mu}+ia sin\theta(\delta_{u}^{\mu}
-\delta_{r}^{\mu})+\frac{i}{sin\theta}\delta_{\phi}^{\mu}),\nonumber\\
\bar{m}^{\mu}&=&\frac{1}{\sqrt{2}r G(r)}(\delta_{\theta}^{\mu}-ia sin\theta(\delta_{u}^{\mu}
-\delta_{r}^{\mu})-\frac{i}{sin\theta}\delta_{\phi}^{\mu}),\label{A7}
\end{eqnarray}
From the definition of the null tetrad, using  Eq. (\ref{A7}) in Eq. (\ref{AA6}), we get a matrix equation.

By using the coefficient of above matrix in Eq. (\ref{AA6}) to obtained $g^{\mu\nu}$ in the EF coordinates are given by

From the definition of the null tetrad the metric tensor $g^{\mu r}$ in the EF coordinate are computed. Furthermore, the matrix's lower index components in the EF coordinates can be expressed as
\begin{eqnarray}
g_{uu}&=&-\frac{a^2 G^2\sigma^2 sin^2\theta+ \tilde{F}G^4\sigma^4}{a^4sin^2\theta},~~~g_{ur}=-\frac{G^2 \sigma^2}{a^2sin^2\theta},~~~
g_{\theta\theta}=\sigma^2G^2,\nonumber\\~~~g_{\phi\phi}&=&-\frac{\tilde{F}G^4\sigma^4}{a^2},~~~
g_{u\phi}=\frac{G^2 \sigma^2(a^2sin^2\theta+\tilde{F}G^2\sigma^2)}{a^3sin^2\theta},~~~
g_{r\phi}=\frac{G^2 \sigma^2}{a},\nonumber
\end{eqnarray}
Finally, we perform the coordinate transformation from the EF to BL coordinates as
\begin{equation}
du=dt+\lambda(r)dr,~~~d\phi=d\phi+h(r)dr\label{B1}
\end{equation}
where 
\begin{eqnarray}
\lambda(r)&=&-\frac{r^2+a^2}{r^2 F +a^2},\nonumber\\
h(r)&=&-\frac{a}{r^2 F +a^2}\nonumber\\
\tilde{F}(r,a,\theta)&=&\frac{r^2 F(r)+a^2 cos^2\theta}{\sigma^2}.\label{B2}
\end{eqnarray}
By using the Eq.(\ref{B1}) and Eq.(\ref{B2}) in Eq. (\ref{A5}) we get the final metric
\begin{equation}
ds^{2}=-F(dt-\lambda(r)dr)^{2}-2(dt-\lambda(r)dr)dr+\sigma^2 d\phi^2 sin^2\theta
+\sigma^2 sin^2\theta( d\phi+h(r)dr)^2.\label{B3}
\end{equation}
From the Eq. (\ref{B3}), we are neglecting non-diagonal values expect $dtd\phi$ as like in these Refs(\cite{R6}-\cite{R9}).
The gauged super gravity like BH metric with new spin (rotation) parameter in $(t, r, \theta, \phi)$ can be derived in the following form
\begin{eqnarray}
ds^{2}&=&-\frac{\sigma^2(a^2 r^2 q_{r} sin^2\theta +k+\frac{\mu}{r^2}g^2 \sigma^2)}{q_{r}}
dt^2+\frac{q_{r}}{\sigma^2(a^2 r^2 q_{r}sin^2\theta+k+\frac{\mu}{r^2}g^2 \sigma^2)}dr^2\nonumber\\
&+&\sigma^2 (\frac{q_{r}}{r^2})d\theta^2
+\frac{r^4 k+\mu r^2 g^2\sigma^4}{q_{r}}d\phi^2 +2\sigma^2(a^2 r^2 sin^2\theta (\frac{q_{r}}{r^2})+k+\frac{\mu}{r^2}g^2 \sigma^2)dtd\phi,\label{m1}
\end{eqnarray}
with $q_{r}=q_{i}+r^2$ and $q_{i}=q_{1}q_{2}q_{3}q_{4}$. Xu et al. calculated \cite{R6} the Kerr-Newman-AdS metric in the Rastall gravity by using the Newman-Janis approach and the rotating metric must satisfy the Einstein-Maxwell field equation. Drake and Turolla discovered \cite{12} that the Newman-Janis method can produce a new Kerr metric interior solution that is suitable. The Kerr metric can be smoothly matched to these new solutions. Since all the necessary supergravity components are now available in equation (\ref{m1}), (gauged) supergravity is a major playground for this modified Janis-Newman algorithm. If the transformation is infinitesimal, it is important to confirm that it is a perfectly valid diffeomorphism and is integrable \cite{1}. Additionally, significant solutions to higher-dimensional Einstein-Maxwell problems remain unsolved (particularly the black rings solution), and it is hoped that an understanding of the Janis-Newman algorithm in these dimensions would help solve this issue. To make the metric simpler, modify the coordinate system (for particular, to the BL system). 
The $T_{H}$ is given as
\begin{equation}
T_{H}=\frac{r^4q_{i}(2r^2+a^2cos^4\theta)\{k+g^2\mu-a^2r^2 cos^2\theta(k+2g^2\mu)-a^4 g^2\mu 
cos^4\theta(q_{i}+2r^2)+r^4 a^2sin^2\theta(q_{i}+2r^2)\}}{2\pi r^3(q_{i}+2r^2)}.\label{T_{H}}
\end{equation}

\section{Corrected tunneling}

The gauge theory that the super-partner of the particle, the gauge boson charged particle, is defined by the super-gravity theory. Additionally, because this theory is more significant than the ungauged case, it contains a negative cosmological constant, where cosmological constant is expressed in an anti de-sitter BH. Now, let's examine the tunnelling of a boson particle from a BH. In the theory of gauged supergravity in four dimensions, we determine the Hawking temperature of BH
bosonic tunneling phenomena at BH horizon. We investigate the stability of charged black holes' thermodynamic properties in gauged supergravity theories with $D=4$. The metric Eq. (\ref{m1}) can be re-written as
\begin{eqnarray}
ds^{2}&=&-Ldt^{2}+\frac{1}{L}dr^{2}+Md\theta^{2}
+N d\phi^{2}+2Zdt d\phi,\nonumber\label{aa}
\end{eqnarray}
where  
\begin{eqnarray}
L&=&\frac{\sigma^2(a^2 r^2 sin^2\theta (q_{i}+r^2)+k+\frac{\mu}{r^2}g^2 \sigma^2)}{q_{i}+r^2},\nonumber\\
M &=& \frac{q_{i}\sigma^2+r^2\sigma^2}{r^2},\nonumber\\
N &=&\frac{r^4 k+\mu r^2 g^2\sigma^4}{q_{i}+r^2},\nonumber\\
\end{eqnarray}
and 
\begin{equation}
Z=2\sigma^2(a^2 r^2 sin^2\theta (\frac{q_{i}+r^2}{r^2})+k+\frac{\mu}{r^2}g^2 \sigma^2).\nonumber
\end{equation}
There are other kinds of solutions, but this review will only focus on four-dimensional BH like solutions, which are described as particle-like boson particles that carry charges like mass. The Lagrangian gravity equation can be defined \cite{4, 22} as  
\begin{eqnarray}
&&\partial_{\mu}(\sqrt{-g}\phi^{\nu\mu})+\sqrt{-g}\frac{m^2}{\hbar^2}\phi^{\nu}+\sqrt{-g}\frac{i}{\hbar}A_{\mu}\phi^{\nu\mu}
+\sqrt{-g}\frac{i}{\hbar}eF^{\nu\mu}\phi_{\mu}+\alpha\hbar^{2}\partial_{0}\partial_{0}\partial_{0}(\sqrt{-g}g^{00}\phi^{0\nu})-
\nonumber\\
&&\alpha \hbar^{2}\partial_{i}\partial_{i}\partial_{i}(\sqrt{-g}g^{ii}\phi^{i\nu})=0,
\end{eqnarray}
here $g$ is determinant coefficient matrix, $\phi^{\nu\mu}$ is anti-symmetric tensor and $m$ is particle mass, since
\begin{eqnarray}
\phi_{\nu\mu}&=&(1-\alpha{\hbar^2\partial_{\nu}^2})\partial_{\nu}\phi_{\mu}-
(1-\alpha{\hbar^2\partial_{\mu}^2})\partial_{\mu}\phi_{\nu}+(1-\alpha{\hbar^2\partial_{\nu}^2})\frac{i}{\hbar}eA_{\nu}\phi_{\mu}
-(1-\alpha{\hbar^2}\partial_{\nu}^2)\frac{i}{\hbar}eA_{\mu}\phi_{\nu},\nonumber\\
&&~~and~~
F_{\nu\mu}=\nabla_{\nu} A_{\mu}-\nabla_{\mu} A_{\nu},\nonumber
\end{eqnarray}
where $\alpha,~A_{\mu},~e~$ and $\nabla_{\mu}$ are the positive dimensionless parameter(GUP parameter),
vector potential of gauge super gravity like BH, the charge of particle and covariant derivatives respectively.
\begin{eqnarray}
&&\phi^{0}=\frac{-N\phi_{0}+Z\phi_{3}}{LN+Z^2},~~~\phi^{1}=L\phi_{1},
~~~\phi^{2}=\frac{1}{M}\phi_{2},~~~
\phi^{3}=\frac{Z\phi_{0}+L\phi_{3}}{LN+Z^2},\nonumber\\
&&\phi^{01}=\frac{(-N\phi_{01}+Z\phi_{13})L}{LN+Z^2},~~~
\phi^{02}=\frac{-N\phi_{02}}{M(LN+Z^2)},
~~~\phi^{03}=\frac{(L^2-LN)\phi_{03}}{(LN+Z^2)^2},\nonumber\\
&&\phi^{12}=\frac{L}{M}\phi_{12},
~\phi^{13}=\frac{L}{LN+Z^2}\phi_{13},~~
\phi^{23}=\frac{Z\phi_{02}+L\phi_{23}}{M(LN+Z^2)},
\end{eqnarray}
The WKB approximation is
\begin{equation}
\phi_{\nu}=c_{\nu}\exp[\frac{i}{\hbar}I_{0}(t,r, \theta, \phi)+
\Sigma \hbar^{n}I_{n}(t,r, \theta, \phi)].
\end{equation}
By computing the Lagrangian equation using the WKB approximation, the set of field equations in \textit{Appendix A} is generated. Using separation of variables technique, we can choose
\begin{equation}
I_{0}=-(E-J\Omega)t+W(r)+J\phi+\nu(\theta),
\end{equation}
where $E$ is the energy of the particle, $J$ represents the particle's angular
momentum corresponding to angles $\phi$.
\begin{eqnarray}
U_{00}&=&-LN\tilde{Z}[W_{1}^2+\alpha W_{1}^4]-\frac{N\tilde{Z}}{M}[J^2+\alpha J^4]
-LN\tilde{Z}^2[\nu_{1}^2+
\alpha \nu_{1}^4]-m^2 N\tilde{Z},\nonumber\\
U_{01}&=&\-LN\tilde{Z}[\tilde{\Omega}+\alpha \tilde{\Omega}^3+eA_{0}+\alpha eA_{0}\tilde{\Omega}W_{1}+LZ\tilde{Z}+[\nu_{1}+
\alpha \nu_{1}^3],\nonumber\\
U_{02}&=&\frac{-N\tilde{Z}}{M}[\tilde{\Omega}+\alpha \tilde{\Omega}^3-eA_{0}-\alpha eA_{0}\tilde{\Omega}^2]J,\nonumber\\
U_{03}&=&-L\tilde{\Omega}\tilde{Z}[W_{1}^2+\alpha W_{1}^4]- \frac{LN\tilde{Z}^{2}}{M}[\tilde{\Omega}+\alpha \tilde{\Omega}^3
-eA_{0}-\alpha eA_{0}\tilde{\Omega}^2]\nu_{1}+m^2Z\tilde{Z},\nonumber\\
U_{10}&=&-LN\tilde{Z}[\tilde{\Omega}W_{1}+\alpha \tilde{\Omega}W_{1}^3]
-m^{2}L-eA_{0}LN\tilde{Z}[W_{1}+\alpha W_{1}^3],\nonumber\\
U_{11}&=&-LN\tilde{Z}[\tilde{\Omega}^2+\alpha \tilde{\Omega}^4-eA_{0}\tilde{\Omega}-\alpha eA_{0}\tilde{\Omega}W_{1}^2]+LZ\tilde{Z}+[\nu_{1}+
\alpha \nu_{1}^3]\tilde{\Omega}-\frac{L}{M}[J^2+\alpha J^4]\nonumber\\&&
-L\tilde{Z}[\nu_{1}
+\alpha \nu_{1}^3]-m^2 L-eA_{0}LN\tilde{Z}
[\tilde{\Omega}+\alpha \tilde{\Omega}^3-eA_{0}-\alpha eA_{0}\tilde{\Omega}^2]
+eA_{0}LZ\tilde{Z}[\nu_{1}+
\alpha \nu_{1}^3],\nonumber\\
U_{12}&=&\frac{L}{M}[W_{1}+\alpha W_{1}^3]J,\nonumber\\
U_{13}&=&-L\tilde{\Omega}\tilde{Z}[W_{1}+\alpha W_{1}^3]\tilde{\Omega}+ L\tilde{Z}^2[W_{1}+\alpha W_{1}^3]\nu_{1}+LZeA_{0}\tilde{Z}[W_{1}+\alpha W_{1}^3],\nonumber
\end{eqnarray}
\begin{eqnarray}
U_{20}&=&\frac{N\tilde{Z}}{M}[\tilde{\Omega}J+\alpha \tilde{\Omega}J^3]+
\frac{Z\tilde{Z}}{M}[\tilde{\Omega}+\alpha \tilde{\Omega}^3\nu_{1}^2]
-\frac{NeA_{0}\tilde{Z}}{M}[J+\alpha J^3],\nonumber\\
U_{21}&=&\frac{L}{M}[J+\alpha J^3]W_{1},\nonumber\\
U_{22}&=&\frac{N\tilde{Z}}{M}[\tilde{\Omega}^2
+\alpha \tilde{\Omega}^4-eA_{0}\tilde{\Omega}-\alpha eA_{0}
\tilde{\Omega}]-\frac{L}{M}+\frac{Z\tilde{Z}}{M}[\tilde{\Omega}
+\alpha \tilde{\Omega}^3-eA_{0}-\alpha eA_{0}\tilde{\Omega}^2]\nu_{1}\nonumber\\&&-\frac{L\tilde{Z}}{M}[\nu_{1}^2+
\alpha \nu_{1}^4]-\frac{m^2}{M}-\frac{eA_{0}N\tilde{Z}}{M}[\tilde{\Omega}+\alpha \tilde{\Omega}^3-eA_{0}-\alpha eA_{0}\tilde{\Omega}^2],\nonumber\\
U_{23}&=&\frac{L\tilde{Z}}{M}[J+\alpha J^3]\nu_{1},\nonumber\\
U_{30}&=&(LN-L^2)\tilde{Z}^2[\nu_{1}+\alpha \nu_{1}^3]\tilde{\Omega}+ \frac{Z\tilde{Z}}{M}[J^2+\alpha J^4]-m^2Z\tilde{Z}-eA_{0}(LN-L^2)\tilde{Z}^2[\nu_{1}+\alpha \nu_{1}^3],\nonumber\\
U_{31}&=&L\tilde{Z}[\nu_{1}+\alpha \nu_{1}^3]W_{1},\nonumber\\
U_{32}&=&\frac{Z\tilde{Z}}{M}[J+\alpha J^3]\tilde{\Omega}+
\frac{L\tilde{Z}}{M}[\nu_{1}+\alpha \nu_{1}^3]J,\nonumber\\
U_{33}&=&(LN-L^2)\tilde{Z}[\tilde{\Omega}^2
+\alpha \tilde{\Omega}^4-eA_{0}\tilde{\Omega}-\alpha eA_{0}\tilde{\Omega}^3]-L\tilde{Z}[W_{1}^2+\alpha W_{1}^4]\nonumber\\&&
-\frac{L\tilde{Z}}{M}[J^2+\alpha J^4]
-m^2 L\tilde{Z}
-eA_{0}(LN-L^2)\tilde{Z}[\tilde{\Omega}
+\alpha \tilde{\Omega}^3-eA_{0}\tilde{\Omega}^2],\nonumber
\end{eqnarray}
where $W_{1}=\partial_{r}{I_{0}}$, $J=\partial_{\phi}I_{0}$, $\nu_{1}=\partial_{\theta}{I_{0}}$,  $\tilde{\Omega}=E-J\Omega$ and $\tilde{Z}=\frac{1}{LN+Z^2}$.
For the non-trivial solution, the determinant $\textbf{U}$ is equal to
zero, so we get
\begin{eqnarray}\label{a1}
ImW^{\pm}_{1}&=&\pm \int\sqrt{\frac{(\tilde{\Omega}-eA_{0})^{2}
+X_{1}[1+\alpha\frac{X_{2}}{X_{1}}]}{\frac{L(LN+Z^2)}{N}}}dr\nonumber\\
&=&\pm i\pi\frac{(\tilde{\Omega}-eA_{0})+[1+\alpha\Xi]}{2\kappa(r_{+})},
\end{eqnarray}
with $\Xi$ is a arbitrary parameter and
\begin{eqnarray}
X_{1}&=&\frac{Z\tilde{Z}}{L)}[\tilde{\Omega}
-eA_{0}]\nu_{1}+\tilde{Z}[\nu_{1}^2]-\frac{m^2}{L},\nonumber\\
X_{2}&=&\frac{N\tilde{Z}}{L}[\tilde{\Omega}^4
-2eA_{0}\tilde{\Omega}^3+(eA_{0})^2\tilde{\Omega}^2]\nonumber\\
&+&\frac{Z\tilde{Z}}{LM}[\tilde{\Omega}^3-eA_{0}\tilde{\Omega}^2]\nu_{1}
-\nu_{1}^4\tilde{Z}-W_{1}^4.\nonumber
\end{eqnarray}
The modified tunneling probability as of charge bosonic particles as given as
\begin{equation}
\Gamma=
exp[{-2\pi}\frac{(E-J\Omega-eA_{0})}
{\kappa(r_{+})}][1+\alpha\Xi].
\end{equation}
The corrected $\acute{T_{H}}$ of BH is given as
\begin{eqnarray}\label{1}
\acute{T_{H}}&=&\frac{[r^4q_{i}(2r^2+a^2cos^4\theta)\{k+g^2\mu-a^2r^2 cos^2\theta(k+2g^2\mu)-a^4 g^2\mu cos^4\theta(q_{i}+2r^2)+r^4 a^2sin^2\theta(q_{i}+2r^2)\}]}{2\pi r^3(q_{i}+2r^2)}\nonumber\\&&[1-\alpha\Xi],
\end{eqnarray}
The modified $\acute{T_{H}}$ of BH depend on metric parameter, quantum gravity parameter as well as rotation parameter. Moreover the Bekenstein entropy for this geometry is defined as
\begin{eqnarray}\label{2}
S=2\pi r_+^2 \sqrt{q_i+r_+^2}.
\end{eqnarray}
\section{Graphical analysis}
In this section, We examine the graphical behaviour of temperature $\acute{T_{H}}$ versus horizon $r_{+}$ over a range of values for the charge $q_{i}$ and GUP parameter $\alpha$.
Only the physical example accurately depicts the stable condition of gauge super gravity like BH. The $\acute{T_{H}}$ strongly increases with the decreasing outer horizon $r +$ and a small parameter value $\Xi=1$ can generate a small difference in temperature. Three distinct conditions are expressed by the starting mass, which really is greater than the remnant mass, depending on different values of the horizon radius, charge, and GUP parameter.
We note that the temperature of the gauge super gravity like BH quantum gravity of 4 dimensions reduces and implies with decreasing GUP parameter.
\begin{center}
\includegraphics[width=8cm]{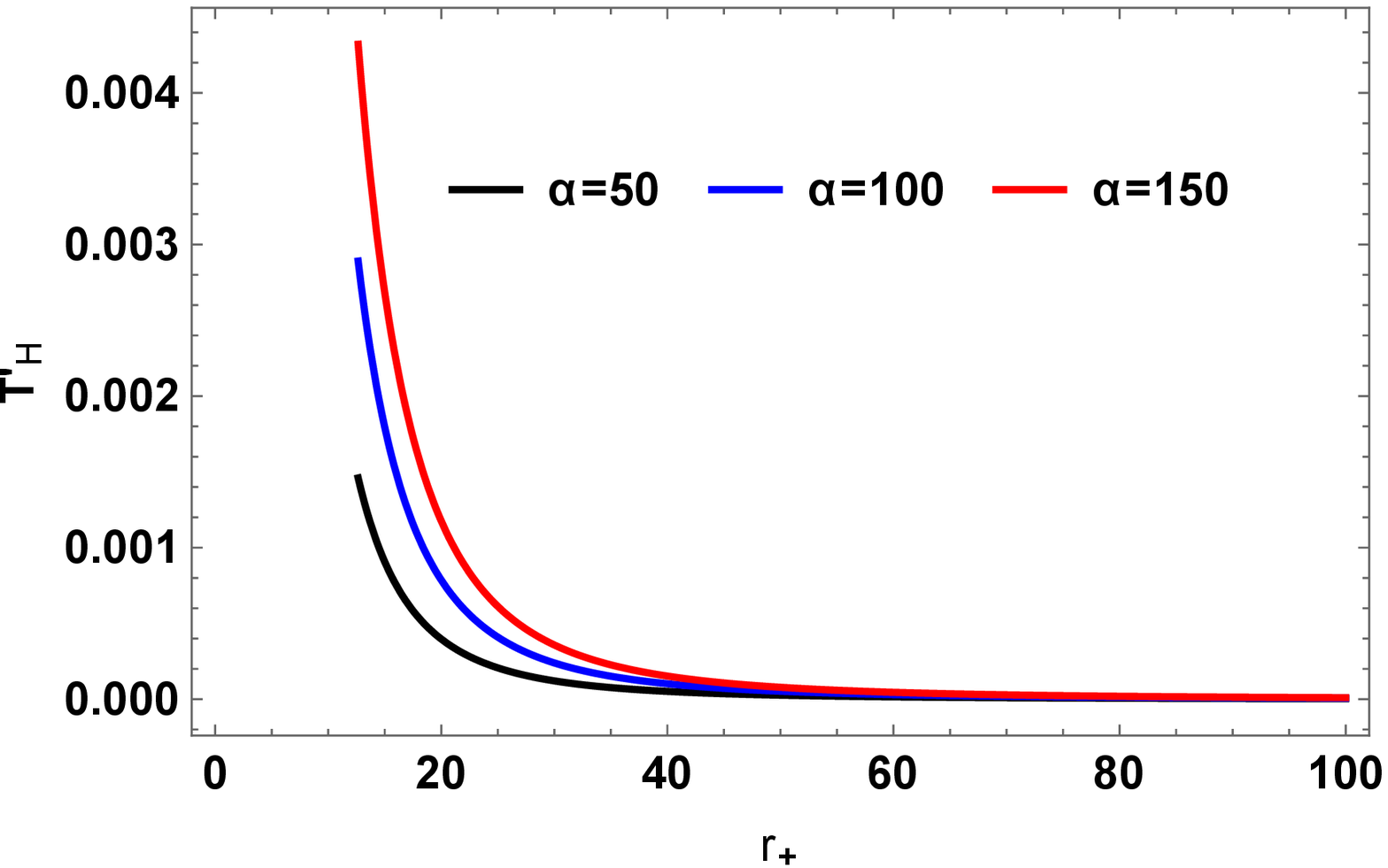}\\
{Figure 1: $\acute{T_{H}}$ versus $r_{+}$ for $\mu=-1,~a=0.3,~k=0.2,~q_{i}=10,~g=0.4,~\Xi=1$ and $\alpha=50~to~150.$ ($i. e:~ i=1~to~4$)}
\end{center}

\section{Thermal Fluctuations}
This section investigates the effects of thermal fluctuations on thermodynamic quantities of gauge super gravity such as BH. We compute the equations of thermodynamics quantities under the logarithmic corrections and graph them for the parameters under consideration. To compute corrected entropy, we first define the partition function as follows.
\begin{equation}\label{19}
R(\varrho)=\int_{0}^{\infty} \exp(-\varrho
\lambda)\rho(\lambda)d\lambda ,
\end{equation}
where $\lambda$ and $\rho(\lambda)$ represent state's average energy and density, respectively. We use the inverse Laplace transform of the partition function defined as to compute $\rho(\lambda)$.
\begin{equation}\label{20}
\rho(\lambda)=\frac{1}{2i\pi
}\int_{-i\infty+\varrho_{0}}^{i\infty+\varrho_{0}}\exp(\varrho
\lambda) R(\varrho) d\varrho =\frac{1}{2i\pi
}\int_{-i\infty+\varrho_{0}}^{i\infty+\varrho_{0}}
\exp(\tilde{S}(\varrho))d\varrho,
\end{equation}
where $\tilde{S}(\varrho)=\beta \lambda+\ln Z(\varrho)$ with $\varrho>0$ depicts the corrected entropy of the BH, which depends on Hawking temperature. The expression of Taylor series for corrected entropy described as
\begin{equation}\label{21}
\tilde{S}(\varrho)=S+\frac{1}{2}(\varrho-\varrho_{0})^{2}
\frac{\partial^{2}\tilde{S}(\varrho)}{\partial
\varrho^{2}}\Big|_{\varrho=\varrho_{0}}+ O(\varrho-\varrho_{o})^{2},
\end{equation}
putting the Eq.(\ref{21}) in (\ref{20}), the equation leads to
\begin{equation}\label{23}
\rho(\lambda)=\frac{1}{\sqrt{2\pi}}\exp(S)\Big(\Big(\frac{\partial^{2}\tilde{S}(\varrho)
}{\partial
\varrho^{2}}\Big)\Big|_{\varrho=\varrho_{0}}\Big)^{-\frac{1}{2}},
\end{equation}
it can be simplified as
\begin{equation}\label{24}
\tilde{S}=S-\frac{1}{2}(\ln S+\ln T^{2})+\frac{\zeta}{S}.
\end{equation}
The number $\frac{1}{2}$ might be used in place of $\omega$ to enhance the the influence of corrections terms. So, the expression of corrected entropy is given as
\begin{equation}\label{25}
\tilde{S}=S-\omega (\ln S+\ln T^{2})+\frac{\zeta}{S}.
\end{equation}
Under first-order corrections, the corrected entropy relation is modified.
The entropy expression is expressed as by ignoring higher order terms.
\begin{equation}\label{25a}
\tilde{S}=S-\omega (\ln S+\ln T^{2}).
\end{equation}
We use Eqs.(\ref{1}) and (\ref{2}) into (\ref{25a}), this leads to
\begin{eqnarray}\label{26}
\tilde{S}&=&2 \pi  r_+^2 \sqrt{q_i+r_+^2}-\omega  \log \Big(\Big(r^3 q_i \sqrt{q_i+r^2} \Big(a^2 \cos^4\theta+2 r_+^2\Big) \Big(a^2 \Big(\Big(q+2 r^2\Big) \Big(r_+^4
   \sin^2\theta- a^2 g^2 \mu  \cos^4\theta\Big)\nonumber\\&-& r_+^2 \cos^2\theta\Big(2 g^2 \mu +k\Big)\Big)+g^2 \mu +k\Big)\Big)\Big(q+2 r_+^2\Big)^{-1}\Big).
\end{eqnarray}
\begin{center}
\includegraphics[width=8cm]{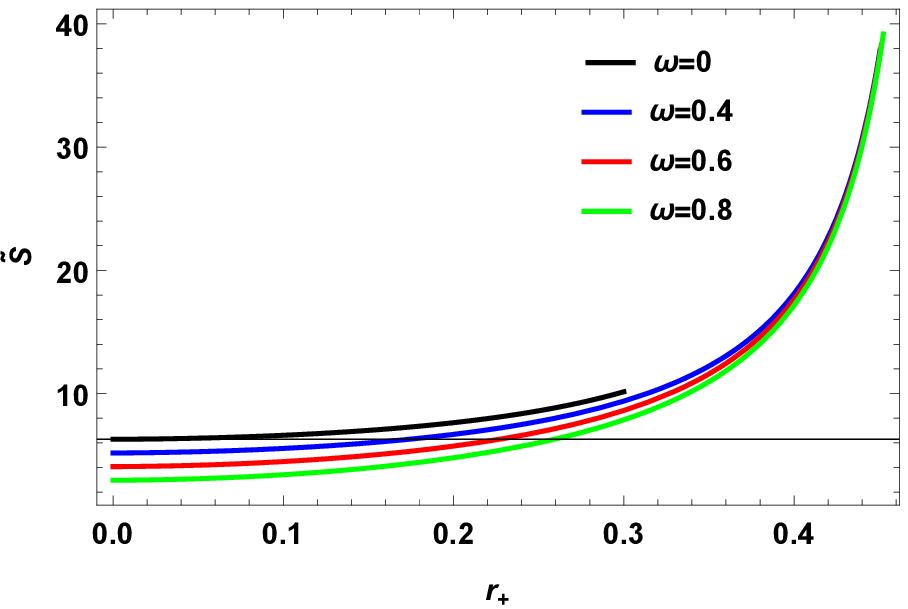}\\
{Figure 3: $\tilde{S}$ versus $r_{+}$ for $\mu$=0.4, g=0.4 and $q_i$=0.2.}
\\
\end{center}
The graphical analysis of corrected entropy for various values of $\omega$ is shown in figure \textbf{3}. It is noted that corrected entropy grows with increasing horizon radius and has a positive trend for all parameter values. It should be emphasized that the second law of thermodynamics is obeyed for the parameter values under consideration, and that both parameters have a minor impact on corrected entropy. Now, using the corrected entropy expression, examine the other quantities of thermodynamic in the presence of thermal fluctuations. So, the energy of Helmholtz ($F=-\int \tilde{S}dT$) goes to the form.
\begin{eqnarray}
F&=&-\Big(r_+^2 q_i \sqrt{q_i+r_+^2} \Big(a^2 \Big(r_+^4 \Big(q+2 r_+^2\Big){}^2 \Big(- \sin^2\theta \Big) \Big(5 a^2  \cos^4\theta+14 r_+^2\Big)+r_+^2
   \cos^2\theta \Big(2 g^2 \mu +k\Big) \Big(a^2  \cos^4\theta \Big(3 q+2 r_+^2\Big)\nonumber\\&+&2 r_+^2 \Big(5 q+6 r_+^2\Big)\Big)+ \cos^4\theta
   \Big(a^4 g^2 \mu  \cos^4\theta \Big(q+2 r_+^2\Big)^2+g^2 \mu  \Big(2 r_+^2 \Big(3 a^2 q^2+1\Big)+24 a^2 q r_+^4+24 a^2 r_+^6-q\Big)-k \nonumber\\&\times&\Big(q-2
   r_+^2\Big)\Big)\Big)-2 r_+^2 \Big(3 q+2 r_+^2\Big) \Big(g^2 \mu +k\Big)\Big)\Big)\Big(\Big(q+2 r_+^2\Big){}^2\Big)^{-1}.
\end{eqnarray}
 \begin{center}
\includegraphics[width=8cm]{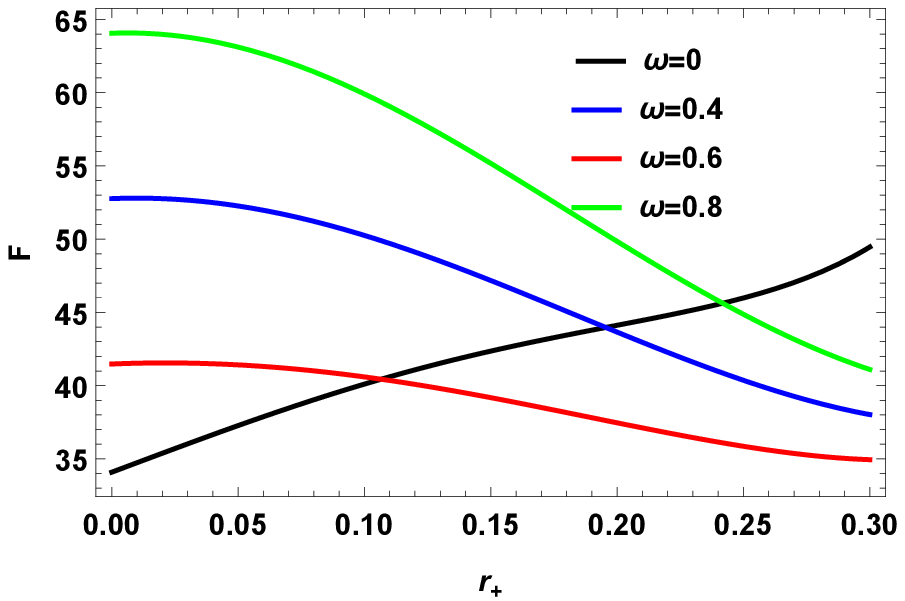}\\
{Figure 3: $F$ versus $r_{+}$ for $\mu$=0.4, g=0.4 and $q_i$=0.2.}
\\
\end{center}
The curve of free energy of Helmholtz w.r.t horizon radius is shown in  \textbf{Fig. 3}. The behavior of energy is seen to steadily decrease for the various correction parameter $\omega$ values. While the uncorrected entropy graph exhibits the opposite graph increases behavior. This action indicates that the system under consideration is shifting toward equilibrium, making it unable to extract any more work from it. For the considered system, the formula for internal energy ($E=F+T\tilde{S}$) is calculated as \cite{z7}  
\begin{eqnarray}
E&=&\Big(r_+^2 q_i \sqrt{q_i+r_+^2} \Big(a^2 \Big(-a^4 g^2 \mu  \Big(r_++1\Big) \cos^4\theta \Big(q+2 r_+^2\Big){}^2+\cos^4\theta \Big(g^2 \mu 
   \Big(-2 a^2 q^2 \Big(r_++3\Big) r_+^2+q\nonumber\\&\times& \Big(-8 a^2 r_+^5-24 a^2 r_+^4+r_++1\Big)-2 r_+^2 \Big(r_+ \Big(4 a^2 r_+^3 \Big(r_++3\Big)-1\Big)+1\Big)\Big)+a^2
   (r+5) r_+^4 \Big(q+2 r_+^2\Big){}^2 \sin^2\theta\nonumber\\&+&k \Big(q \Big(r_++1\Big)+2 \Big(r_+-1\Big) r_+^2\Big)\Big)+r_+^2 \cos^2\theta \Big(2 g^2
   \mu +k\Big) \Big(a^2 \Big(-\cos^4\theta\Big) \Big(q \Big(r_++3\Big)+2 \Big(r_++1\Big) r_+^2\Big)\nonumber\\&-&2 r_+^2 \Big(q \Big(r_++5\Big)+2
   \Big(r_++3\Big) r_+^2\Big)\Big)+2 \Big(r_++7\Big) r_+^6 \Big(q+2 r_+^2\Big){}^2 \sin^2\theta\Big)+2 r_+^2 \Big(q \Big(r_++3\Big)+2
   \Big(r_++1\Big) r_+^2\Big) \nonumber\\&\times&\Big(g^2 \mu +k\Big)\Big)\Big)\Big(\Big(q+2 r_+^2\Big){}^2\Big)^{-1}.
\end{eqnarray}
\begin{center}
\includegraphics[width=8cm]{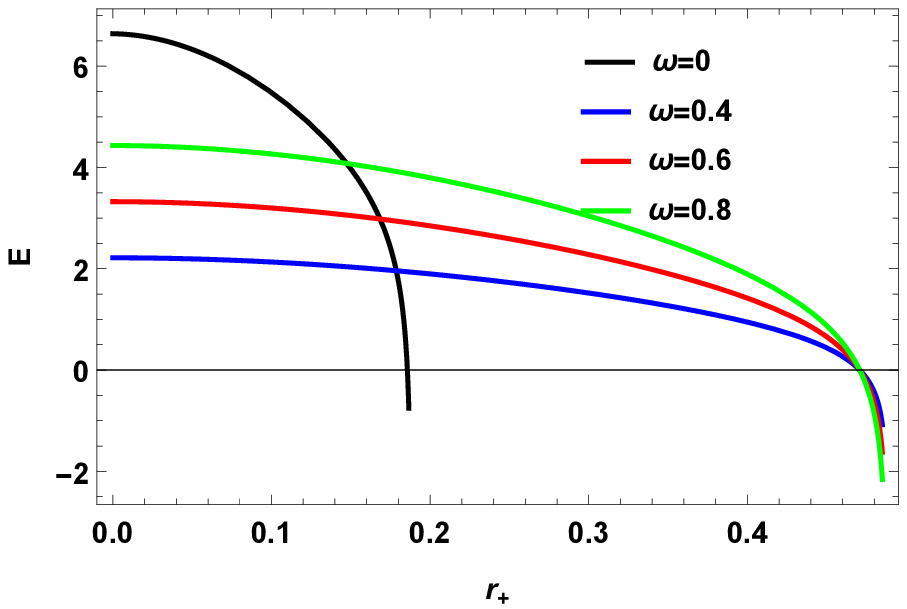}\\
{Figure 4: $E$ versus $r_{+}$ for $\mu$=0.4, g=0.4 and $q_i$=0.2.}
\\
\end{center}
In \textbf{Fig. 4}, the internal energy behavior for the various horizon radius options is depicted graphically. It should be noticed that the graph rapidly lowers and even shifts to the negative side for tiny values of radii, but the $E$ shows physical behavior. This implies that in order to retain its state, the considered BH must absorb an increasing amount of heat from its surroundings. Since BHs are thought to as thermodynamic systems, pressure is another crucial thermodynamic quantity. The form of the BH pressure ($P=-\frac{dF}{dV}$) expression when thermal fluctuations are present is
\begin{eqnarray}
P&=&\Big(r_+ q_i \Big(a^6 g^2 \mu  \cos^4\theta \Big(q+2 r_+^2\Big){}^3 \Big(2 q_i+3 r_+^2\Big)+a^2 \cos^4\theta \Big(2 q_i \Big(g^2 \mu  \Big(12
   a^2 q^3 r_+^2+q^2 \Big(72 a^2 r_+^4-1\Big)+6 q \nonumber\\&\times& \Big(24 a^2 r_+^6+r_+^2\Big)+96 a^2 r_+^8\Big)-k q \Big(q-6 r_+^2\Big)\Big)+r_+^2 \Big(g^2 \mu  \Big(4 r_+^4
   \Big(45 a^2 q^2+1\Big)+6 qr_+^2 \Big(5 a^2 q^2+2\Big)+360 a^2 q r_+^6 \nonumber\\&+& 240 a^2 r_+^8-3 q^2\Big)+k \Big(-3 q^2+12 q r_+^2+4 r_+^4\Big)\Big)-5 a^2 r_+^4 \Big(q+2
   r_+^2\Big){}^3 \sin^2\theta\Big(6 q_i+7 r_+^2\Big)\Big)+a^2 r_+^2 \cos^2\theta\Big(2 g^2 \mu +k\Big) \nonumber\\&\times&\Big(a^2 \cos^2\theta \Big(4
   q_i \Big(3 q^2+3 q r_+^2+2 r_+^4\Big)+15 q^2 r_+^2+20 q r_+^4+12 r_+^6\Big)+2 \Big(q_i \Big(30 q^2 r_+^2+68 q r_+^4+48 r_+^6\Big)+35 q^2 r_+^4+84 q r_+^6\nonumber\\&+&60
   r_+^8\Big)\Big)-2 r_+^2 \Big(7 a^2 r_+^4 \Big(q+2 r_+^2\Big){}^3 \sin^2\theta \Big(8 q_i+9 r_+^2\Big)+\Big(g^2 \mu +k\Big) \Big(4 q_i \Big(3 q^2+3 q
   r_+^2+2 r_+^4\Big)+15 q^2 r_+^2+20 q r_+^4\nonumber\\&+&12 r_+^6\Big)\Big)\Big)\Big)\Big(\Big(q+2 r_+^2\Big){}^3 \sqrt{q_i+r_+^2}\Big)^{-1}.
   \end{eqnarray}
 \begin{center}
\includegraphics[width=8cm]{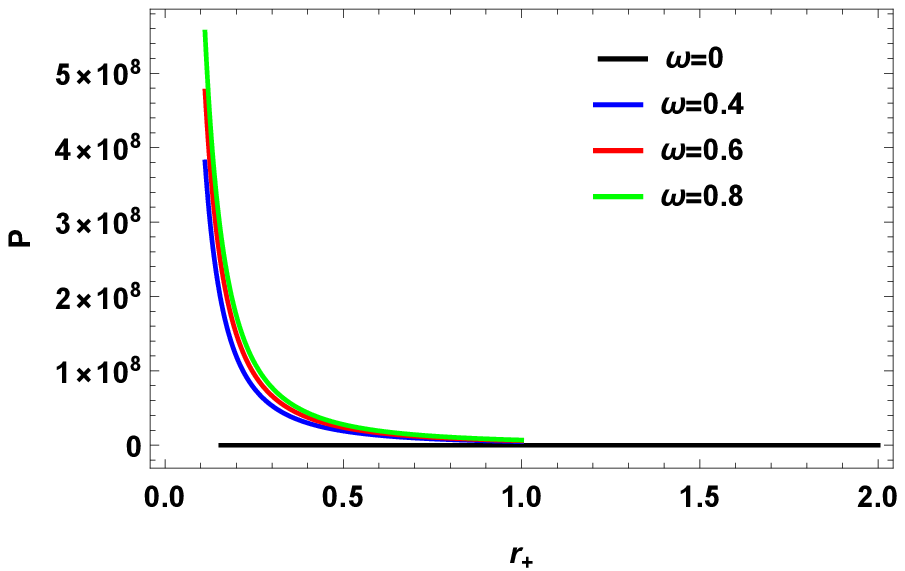}\\
{Figure 5: $P$ versus $r_{+}$ for $\mu$=0.4, g=0.4 and $q_i$=0.2.}
\\
\end{center}
The pressure graph in \textbf{Fig. 5} is coincides with the point of equilibrium. For various correction parameter values, the pressure dramatically rises for the system under consideration. Enthalpy ($H=E+PV$) is a further significant thermodynamic quantity that is given in \textbf{Appendix B}.
  \begin{center}
\includegraphics[width=8cm]{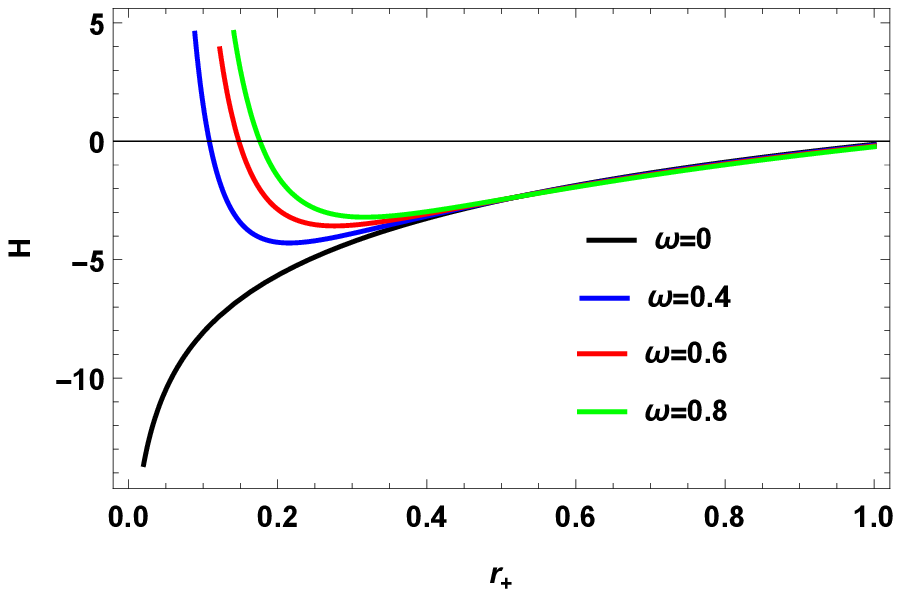}\\
{Figure 6: $H$ versus $r_{+}$ for $\mu$=0.4, g=0.4 and $q_i$=0.2.}
\\
\end{center}
\textbf{Fig. 6} shows that the corrected enthalpy views with the corrected one plot and abruptly shifts to the side of negative. This implies that reactions of exothermic exist and that a significant amount of energy will be released into the environment. The Gibbs free energy ($G=H-T\tilde{S}$) under the effect of thermal fluctuations is stated in \textbf{Appendix B}.
     \begin{center}
\includegraphics[width=8cm]{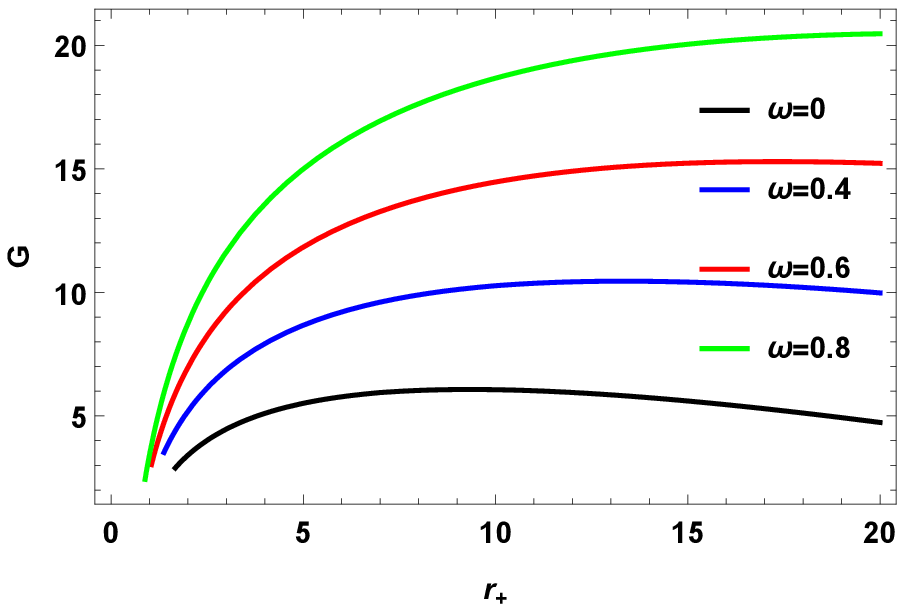}\\
{Figure 7: $G$ versus $r_{+}$ for $\mu$=0.4, g=0.4 and $q_i$=0.2.}
\\
\end{center}
From \textbf{Fig. 7} illustrates the $G$ w.r.t the horizon radius. Positive energy indicates the non-spontaneous presence processes, which is the system to reach equilibrium needs more energy. The stability of the system, which is verified by specific heat ($C_{\tilde{S}}=\frac{dE}{dT}$) is given in \textbf{Appendix B}.
  \begin{center}
  \includegraphics[width=8cm]{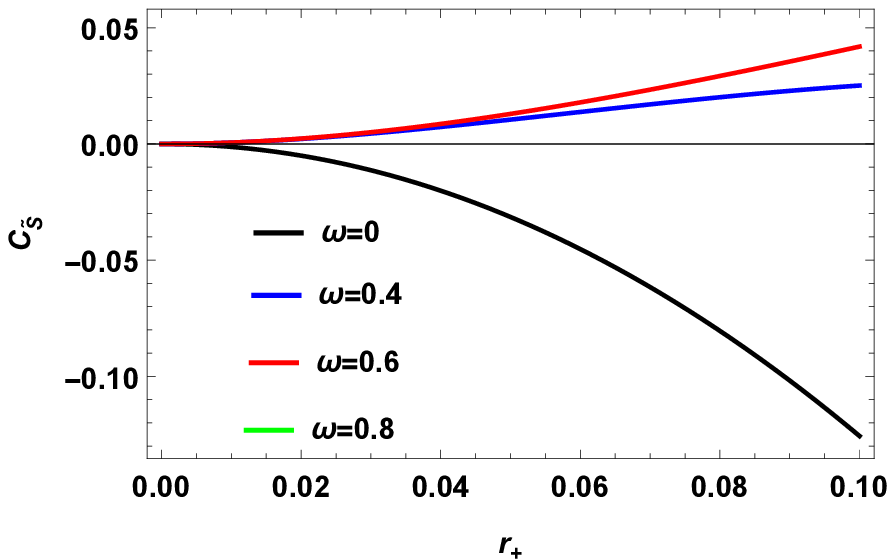}\\
{Figure 8: $C_{\tilde{S}}$ versus $r_{+}$ for $\mu$=0.4, g=0.4 and $q_i$=0.2.}
\\
\end{center}
In the \textbf{Fig. 8}, the behavior of $C_{\tilde{S}}$ is seen in relation to the horizon radius under the various $\omega$ selections. It is clear that whereas the quantity of uncorrected represents negative presentation, indicating that the unstable system, the $C_{\tilde{S}}$ exhibits positive presentation across the whole examined area. This plot's positivity is a sign of a stable area. It is evident that the thermal fluctuations under the correction terms arrives at the stable of system. 
\section{Discussion}
In this paper, 
the quantum tunneling mechanism of massive boson particles from a gauge super gravity like BH have examined. First, the particles motion deal field equation with GUP parameter by using the quantum tunneling approach and a $(t,r,\theta,\phi)$ kind of coordinate system.
To determine the modified tunneling rate of the bosonic particles, we examined the semi-classical action in Planck's constant power series. Then, using this methodology appropriately, the surface gravity has estimated while neglecting the back reaction influences of space-time. We reached the realization that the modified tunneling probability depends not only on the features of BHs and GUP parameter but also on the properties of the emitted vector.

We investigated the thermodynamic quantities heat capacity, Hawking temperature, and corrected entropy in gauge super gravity systems. For stability reasons, we first derived the expressions for entropy, Hawking temperature, and heat capacity, and then calculated the state's density using the inverse Laplace transformation. We investigated various thermodynamic quantities such as Helmholtz free energy, internal energy, Gibbs free energy, enthalpy, pressure, and heat capacity using corrected entropy and also checked stability of the system by two different techniques such as heat capacity and Hessian matrix trace. Under the influence of first-order corrections, we plot a corrected entropy graph for $\omega$ parameter values. Throughout the domain under consideration, the graph of corrected entropy increases monotonically. The graph (black) of usual entropy increases only for small values of horizon radius, whereas the corrected expression of energy increases smoothly. As a result, for small BHs, these correction terms are more effective. Now, using the corrected entropy expression, examine the other thermodynamic quantities in the presence of thermal fluctuations. 

Furthermore, for different correction parameter $\omega$ values, the behavior of Helmholtz energy gradually decreases. The entropy, on the other hand, exhibits the graph increases in the other direction. This observation indicates that the studied system is approaching equilibrium and no further work can be produced from it. The graphical behavior of internal energy is also examined for various horizon radius choices. It is worth noting that for small radii values, while the studied internal energy displays positive behaviour, the graph steadily declines and even shifts to the negative side. The plots of the corrected enthalpy graph match up with the standard enthalpy graph, which then sharply declines and shifts to the negative side. As a result, there will be an exothermic reaction, and a significant quantity of energy will be released into the environment. Furthermore, \textbf{Fig. 7} depicts the Gibbs free energy in terms of the horizon radius. Non-spontaneous reactions suggest that more energy is needed for this system to reach equilibrium. \textbf{Fig. 8} illustrates how specific heat behaves in relation to horizon radius and various correction parameter options.

{\bf Acknowledgement}\\
 PKS acknowledges the National Board for Higher Mathematics (NBHM) under the Department of Atomic Energy (DAE), Govt. of India for financial support to carry out the Research project No.: 02011/3/2022 NBHM(R.P.)/R \& D II/2152 Dt.14.02.2022. We are very much grateful to the honorable referees and to the editor for the illuminating suggestions that have significantly improved our work in terms of research quality, and presentation.
\section*{Appendix A}
By computing the Lagrangian equation using the WKB approximation, the set of field equations is generated as
\begin{eqnarray}
&&\frac{LN}{N+Z^2}[c_{1}(\partial_{0}I_{0})(\partial_{1}I_{0})+\alpha c_{1}
(\partial_{0}I_{0})^{3}(\partial_{1}I_{0})-c_{0}(\partial_{1}I_{0})^{2}
- c_{0}(\partial_{1}I_{0})^4\alpha+eA_{0}(\partial_{1}I_{0})c_{1}+\alpha eA_{0}(\partial_{0}I_{0})^{2}(\partial_{1}I_{0})c_{1}]
\nonumber\\
&&-\frac{LZ}{LN+Z^2}[c_{3}(\partial_{1}I_{0})^2+\alpha c_{3}(\partial_{1}I_{0})^4
-c_{1}(\partial_{1}I_{0})(\partial_{3}I_{0})-\alpha c_{1}(\partial_{1}I_{0})(\partial_{3}I_{0})^2]
+\frac{N}{M(LN+Z^2)}[c_{2}(\partial_{0}I_{0})(\partial_{2}I_{0})
+\alpha c_{2}\nonumber\\
&&(\partial_{0}I_{0})^3(\partial_{2}I_{0})
-c_{0}(\partial_{2}I_{0})^2-\alpha c_{0}(\partial_{2}I_{0})^4+c_{2}eA_{0}(\partial_{2}I_{0})+c_{2}eA_{0}\alpha(\partial_{0}I_{0})^{2}(\partial_{1}I_{0})]+
\frac{LN}{(LN+Z^2)^2}[c_{3}(\partial_{0}I_{0})(\partial_{3}I_{0})
\nonumber\\
&&+\alpha c_{3}(\partial_{0}I_{0})^{3}(\partial_{3}I_{0})-c_{0}(\partial_{3}I_{0})^{2}
-\alpha c_{0}(\partial_{3}I_{0})^4
+c_{3}eA_{0}(\partial_{3}I_{0})+c_{3}eA_{0}(\partial_{0}I_{0})^{2}(\partial_{3}I_{0})]-m^2\frac{N c_{0}-Z c_{3}}{LN+Z^2}=0,\\
&&\frac{-LN}{LN+Z^2}[c_{1}(\partial_{0}I_{0})^2+\alpha c_{1}
(\partial_{0}I_{0})^4-c_{0}(\partial_{0}I_{0})(\partial_{1}I_{0})-\alpha c_{0}(\partial_{0}I_{0})(\partial_{1}I_{0})^{3}
+c_{1}eA_{0}(\partial_{0}I_{0})+\alpha c_{1}eA_{0}(\partial_{0}I_{0})^3]+\nonumber\\
&&\frac{LZ}{LN+Z^2}
[c_{3}(\partial_{0}I_{0})(\partial_{1}I_{0})+\alpha c_{3}
(\partial_{0}I_{0})(\partial_{1}I_{0})^3-c_{1}(\partial_{0}I_{0})(\partial_{3}I_{0})-\alpha c_{1}(\partial_{0}I_{0})(\partial_{3}I_{0})^{3}]+\frac{L}{M}[c_{2}(\partial_{1}I_{0})(\partial_{2}I_{0})+\alpha c_{2}\nonumber\\
&&(\partial_{1}I_{0})(\partial_{2}I_{0})^3-c_{1}(\partial_{2}I_{0})^{2}-\alpha c_{1}(\partial_{2}I_{0})^{4}]+\frac{L}{LN+Z^2}[c_{3}(\partial_{1}I_{0})(\partial_{3}I_{0})
+\alpha c_{3}(\partial_{1}I_{0})(\partial_{3}I_{0})^3-c_{1}(\partial_{3}I_{0})^2-\alpha c_{1} (\partial_{3}I_{0})^{4}]\nonumber\\
&&-m^2 Lc_{1}
+\frac{eA_{0}LN}{LN+Z^2}[c_{1}(\partial_{0}I_{0})+\alpha c_{1}(\partial_{0}I_{0})^3
-c_{0}(\partial_{1}I_{0})-\alpha c_{0}(\partial_{1}I_{0})^3+eA_{0}c_{1}+\alpha c_{1}eA_{0}(\partial_{0}I_{0})^{2})]+\nonumber\\
&&\frac{eA_{0}LZ}{LN+Z^2}[c_{3}(\partial_{1}I_{0})+\alpha c_{3}(\partial_{1}I_{0})^3
-c_{1}(\partial_{3}I_{0})-\alpha c_{1}(\partial_{1}I_{0})^3]=0,\\
&&\frac{N}{M(LN+Z^2)}[c_{2}(\partial_{0}I_{0})^2+\alpha c_{2}
(\partial_{0}I_{0})^{4}-c_{0}(\partial_{0}I_{0})(\partial_{2}I_{0})
-\alpha c_{0}(\partial_{0}I_{0})(\partial_{2}I_{0})^3
+c_{2}eA_{0}(\partial_{0}I_{0})+\alpha c_{2}eA_{0}(\partial_{0}I_{0})^{3}]
\nonumber\\
&&+\frac{L}{M}[c_{2}(\partial_{1}I_{0})^2+\alpha c_{2}
(\partial_{1}I_{0})^{4}-c_{1}(\partial_{1}I_{0})(\partial_{2}I_{0})
-\alpha c_{1}(\partial_{1}I_{0})(\partial_{2}I_{0})^3]-\frac{Z}{M(LN
+Z^2)}[c_{2}(\partial_{0}I_{0})(\partial_{3}I_{0})+\alpha c_{2}
(\partial_{0}I_{0})^{3}\nonumber\\
&&(\partial_{3}I_{0})-c_{0}(\partial_{0}I_{0})(\partial_{3}I_{0})
-\alpha c_{0}(\partial_{0}I_{0})^3 (\partial_{3}I_{0})+c_{2}eA_{0}(\partial_{3}I_{0})
+\alpha c_{2}eA_{0}(\partial_{3}I_{0})^{3}]
+\frac{L}{M(LN+Z^2)}[c_{3}(\partial_{2}I_{0})(\partial_{3}I_{0})\nonumber\\
&&+\alpha c_{3}
(\partial_{2}I_{0})^{3}(\partial_{3}I_{0})-c_{2}(\partial_{3}I_{0})^2
-\alpha c_{2}(\partial_{3}I_{0})^4]
-\frac{m^2 c_{2}}{M}+\frac{eA_{0}N}{M(LN+Z^2)}[c_{2}(\partial_{0}I_{0})+\alpha c_{2}(\partial_{0}I_{0})^3-(\partial_{2}I_{0})c_{0}\nonumber\\
&&-(\partial_{2}I_{0})^3c_{0}\alpha+eA_{0}c_{2}+ eA_{0}\alpha(\partial_{0}I_{0})^2c_{2}]=0,\\
&&\frac{LN-L^2}{(LN+Z^2)^2}[(\partial_{0}I_{0})^2c_{3}+\alpha
(\partial_{0}I_{0})^4c_{3}-(\partial_{0}I_{0})(\partial_{3}I_{0})c_{0}-\alpha (\partial_{0}I_{0})(\partial_{3}I_{0})^{3}c_{0}+{eA_{0}c_3}(\partial_{0}I_{0})
+\alpha c_{3}eA_{0}(\partial_{0}I_{0})^{3}]
\nonumber\\
&&-\frac{N}{M(LN+Z^2)}[c_{3}(\partial_{1}I_{0})^2+\alpha c_{3}
(\partial_{1}I_{0})^{4}
-c_{1}(\partial_{1}I_{0})(\partial_{3}I_{0})
-\alpha c_{1}(\partial_{1}I_{0})(\partial_{3}I_{0})^3]-\frac{Z}{M(LN+Z^2)}
[c_{2}(\partial_{0}I_{0})(\partial_{2}I_{0})\nonumber\\
&&+\alpha c_{2}
(\partial_{0}I_{0})^3(\partial_{2}I_{0})-c_{0}(\partial_{2}I_{0})^{2}
+\alpha c_{0}(\partial_{2}I_{0})^4+{eA_{0}c_2}(\partial_{2}I_{0})
+\alpha c_{2}eA_{0}(\partial_{0}I_{0})^{2}(\partial_{2}I_{0})]-\frac{eA_{0}L}{M(LN+Z^2)}
\nonumber\\
&&[c_{3}(\partial_{2}I_{0})^2+\alpha c_{3}(\partial_{2}I_{0})^4-c_{2}(\partial_{2}I_{0})(\partial_{3}I_{0})
-\alpha c_{2}(\partial_{0}I_{0})(\partial_{3}I_{0})^{3}]
-\frac{m^2 (Zc_{0}-Lc_{3})}{LN+Z^2}
+\frac{eA_{0}(LN-L^2)}{(LN+Z^2)^2}[c_{3}(\partial_{0}I_{0})+\alpha c_{3}\nonumber\\
&&(\partial_{0}I_{0})^3
-(\partial_{3}I_{0})c_{0}-\alpha (\partial_{3}I_{0})^3c_{0}+eA_{0}c_{3}
+\alpha eA_{0}(\partial_{0}I_{0})^2c_{3}]=0.
\end{eqnarray}
\section*{Appendix B}
Enthalpy is computed as
\begin{eqnarray}
H&=&(r_+ q_i (4 r_+ (2 r_+^2+q) ((2 \sin^2\theta  (r_++7) (2 r_+^2+q)^2 r_+^6+(2 \mu  g^2+k)
  \cos^2\theta  (-\cos^4\theta  (2 (r_++1) r_+^2\nonumber\\&&+q (r_++3)) a^2-2 r_+^2(2 (r_++3) r_+^2+q
   (r_++5))) r_+^2-a^4 g^2 \mu  \cos^4\theta  (r_++1) (2 r_+^2+q){}^2+ \nonumber\\&&\cos^4\theta  (a^2 \sin^2\theta(r_++5) (2 r_+^2+q)^2 r_+^4+k (2 (r_+-1) r_+^2+q (r_++1))+g^2 \mu  (-2 a^2 q^2 (r_++3) r_+^2 \nonumber\\&&-2
   (r_+ (4 a^2 r_+^3 (r_++3))+1) r_+^2+q (-8 a^2 r_+^5-24 a^2 r_+^4+r_++1)))) a^2+2 (\mu  g^2+k)
   r_+^2 \nonumber\\&&(2 (r_++1) r_+^2 +q (r_++3))) (r_+^2+q_i)-(3 g^2 (g^2 \mu\cos^4\theta  (2
   r_+^2+q)^3 (3 r_+^2+2 q_i) a^6+\cos^4\theta \nonumber\\&& (-5 a^2 \sin^2\theta  (2 r_+^2+q)^3 (7 r_+^2+6 q_i)
   r_+^4+(\mu (240 a^2 r_+^8+360 a^2 q r_+^6+4 (45 a^2 q^2+1) r_+^4+6 q (5 a^2 q^2+2) \nonumber\\&& r_+^2-3 q^2) g^2+k (-3 q^2+4
   r_+^2)) r_+^2+2 (g^2 \mu (96 a^2 r_+^8+12 a^2 q^3 r_+^2+q^2 (72 a^2 r_+^4-1)+6 q (24 a^2 r_+^6+r_+^2))\nonumber\\&&-k q (q-6
   r_+^2)) q_i) a^2+(2 \mu  g^2+k) \cos^2\theta r_+^2 (\cos^4\theta (12 r_+^6+20 q r_+^4+15 q^2 r_+^2+4 (2 r_+^4+3
   q r_+^2+3 q^2) q_i) \nonumber\\&& a^2+2 (60 r_+^8+84 q r_+^6+35 q^2 r_+^4+(48 r_+^6+68 q r_+^4+30 q^2 r_+^2) q_i)) a^2-2 r_+^2 (7 a^2
   \sin^2\theta (2 r_+^2+q)^3 \nonumber\\&&(9 r_+^2+8 q_i) r_+^4+(\mu  g^2+k) (12 r_+^6+20 q r_+^4+15 q^2 r_+^2+4 (2 r_+^4+3 q r_+^2+3
   q^2) q_i))))(\pi )^{-1}))\nonumber\\&&(4 (q+2 r_{+}^{2})^3 \sqrt{r_+^2+q_i})^{-1}.
  \end{eqnarray}
The Gibbs free energy is stated as
\begin{eqnarray} 
  G&=&( r_+^4 (\cos^4\theta a^2+2 r_+^2) (((2 r_+^2+q) (\sin^2\theta r_+^4-a^2 g^2 \mu  \cos^4\theta)-r^2
(2 \mu  g^2+k)\cos^2\theta ) a^2+k+g^2 \mu )\nonumber\\&& q_i^2 (4 r_+ (2 r_+^2+q) ((2 \cos^2\theta (r_++7)
   (2 r_+^2+q)^2 r_+^6+(2 \mu  g^2+k) \cos^2\theta (-\cos^4\theta(2 (r_++1) r_+^2+\nonumber\\&& q (r_++3))
   a^2-2 r_+^2 (2 (r_++3) r_+^2+q (r_++5))) r_+^2-a^4 g^2 \mu  \cos^2\theta(r_++1) (2
   r_+^2+q){}^2+\cos^4\theta\nonumber\\&& (a^2 \sin^2\theta (r_++5) (2 r_+^2+q)^2+k (2 (r_+-1) r_+^2+q
   (r_++1))+g^2 \mu (-2 a^2 q^2 (r_++3) r_+^2-2\nonumber\\&& (r_+ (4 a^2 r_+^3 (r_++3)-1)+1) r_+^2+q (-8 a^2 r_+^5-24
   a^2 r_+^4+r_++1)))) a^2+2 (\mu  g^2+k) r_+^2 (2 (r_++1) \nonumber\\&& r_+^2+q (r_++3)))
   (r_+^2+q_i)-(3 g^2 (g^2 \mu  \cos^4\theta(2 r_+^2+q)^3 (3 r_+^2+2 q_i) a^6+\cos^4\theta (-5 a^2
  \sin^2\theta \nonumber\\&& (2 r_+^2+q)^3 (7 r_+^2+6 q_i) r_+^4+(\mu(240 a^2 r_+^8+360 a^2 q r_+^6+4 (45 a^2 q^2+1) r_+^4+6 q(5 a^2 q^2+2) r_+^2-3 q^2) g^2\nonumber\\&&+k (4 r_+^4+12 q r_+^2-3 q^2)) r_+^2+2 (g^2 \mu (96 a^2 r_+^8+12 a^2 q^3 r_+^2+q^2 (72 a^2r_+^4-1)+6 q (24 a^2 r_+^6+r_+^2))-k q \nonumber\\&& (q-6 r_+^2)) q_i) a^2+(2 \mu  g^2+k) \cos^2\theta r_+^2
  (\cos^4\theta (12 r_+^6+20 q r_+^4+15 q^2 r_+^2+4 (2 r_+^4+3 q r_+^2+3 q^2) q_i) a^2\nonumber\\&&+2 (60 r_+^8+84 q r_+^6+35 q^2 r_+^4+(48
   r_+^6+68 q r_+^4+30 q^2 r_+^2) q_i)) a^2-2 r_+^2 (7 a^2 \sin^2\theta(2 r_+^2+q)^3 \nonumber\\&&(9 r_+^2+8 q_i) r_+^4+(\mu
   g^2+k) (12 r_+^6+20 q r_+^4+15 q^2 r_+^2+4 (2 r_+^4+3 q r_+^2+3 q^2) q_i))))(\pi)^{-1}))\nonumber\\&&(4 (q)^4)^{-1}.
   \end{eqnarray}
The specific heat is given as
\begin{eqnarray}
  C_{\tilde{S} }&=&-(2 \text{$\pi $r}_+ (-g^2 \mu \cos^4\theta (2 r_+^2+q)^3 ((4 r_++3) r_+^2+(3 r_++2) q_i)
   a^6+ \cos^4\theta(a^2  \sin^2\theta (2 r_+^2+q)^3 \nonumber\\&&((8 r_++35) r_+^2+(7 r_++30) q_i) r_+^4+(k(4
   (2 r_+-1) r_+^4+12 q (r_+-1) r_+^2+q^2 (4 r_++3))-g^2 \mu  \nonumber\\&&(4(24 a^2 r_+^5+60 a^2 r_+^4-2 r_++1) r_+^4+6 a^2 q^3
   (2 r_++5) r_+^2+12 q (r_+ (6 a^2 r_+^3 (2 r_++5)-1)+1) r_+^2\nonumber\\&&+q^2 (4 r_+ (9 a^2 r_+^3 (2
   r_++5)-1)-3))) r_+^2+(k (4 r_+^5+4 q (2 r_+-3) r_+^2+q^2 (3 r_++2))-g^2 \mu \nonumber\\&& (4 (4 a^2 r_+^3
   (5 r_++12)-1) r_+^5+2 a^2 q^3 (5 r_++12) r_+^2+4 q (30 a^2 r_+^5+72 a^2 r_+^4-2 r_+^2+3) r_+^2+q^2 \nonumber\\&&(3 r_+ (4 a^2 r_+^3
   (5 r_++12)-1)-2))) q_i)-(2 \mu  g^2+k)  \cos^2\theta r_+^2 (2 (12 (2 r_+^2+5) r_+^4+28 q
   (r_++3) \nonumber\\&& r_+^2+q^2 (8 r_++35)) r_+^4+2(4 (5 r_++12) r_+^4+4 q (6 r_++17) r_+^2+q^2 (7 r_++30)) q_i
   r_+^2+a^2  \cos^4\theta \nonumber\\
   &&((4(4 r_++3) r_+^4+20 q (r_++1) r_+^2+3 q^2 (2 r_++5)) r_+^2+(4(3 r_++2)
   r_+^4+ (4 r_++3)+q^2 \nonumber\\&&(5 r_++12)) q_i)) a^2+2 r_+^2 (a^2  \sin^2\theta (2 r_+^2+q)^3
  ((10 r_++63) r_+^2+(9 r_+^2+56)) r_+^4+(\mu  g^2+k)\nonumber\\&& ((4(4 r_++3) r_+^4+20 q (r_++1) r_+^2+3
   q^2 (2 r_+^2+5)) r_+^2+(4(3 r_++2) r_+^4+4 q(4 r_++3) r_+^2\nonumber\\&&+q^2 (5 r_++12)) q_i))))((2
   r_+^2+q)(a^2 (- \sin^2\theta (2 r_+^2+q)^2 \Big(5  \cos^4\theta a^2+14 r_+^2) r_+^4+(2 \mu  g^2+k)\nonumber\\&&
   \cos^2\theta(\cos^4\theta (2 r_+^2+3 q) a^2+2 r_+^2 (6 r_+^2+5 q)) r_+^2+\ \cos^4\theta 
   (g^2 \mu 
    \cos^4\theta (2 r_+^2+q){}^2 a^4-k (q-2 r_+^2)\nonumber\\&&+g^2 \mu (24 a^2 r_+^6+24 a^2 q r_+^4+2 (3 a^2 q^2+1)
   r_+^2-q)))-2 (\mu  g^2+k) r_+^2 (2 r_+^2+3 q)) \sqrt{r_+^2+q_i})^{-1}.
  \end{eqnarray}


\begin{thebibliography}{99}

\bibitem{1} H. Erbin, Universe {\bf 3} (2017) 19.

\bibitem{2}	M. Azreg-A\"{i}nou, Phys. Lett. {\bf B 730} (2014) 95.

\bibitem{2a} R. Ali, R. Babar, M. Asgher and T. C. Xia, Int. J. Mod. Phys. {\bf A 37} (2022) 2250108.

\bibitem{3}	R. Ali, R. Babar and M. Asgher, Annalen der Physik, {\bf 2200074} (2022) 12.

\bibitem{4} R. Ali, R. Babar and P. K. Sahoo, Phys. Dark Universe, {\bf 35} (2022) 100948.

\bibitem{4a} R. Ali, R. Babar, M. Asgher and G. Mustafa, Int. J. Mod. Phys. {\bf A 37} (2022) 2250134.

\bibitem{5} E.T. Newman and A.I. Janis, J. Math. Phys. {\bf 6} (1965) 915.

\bibitem{6} E.T. Newman, et al, J. Math. Phys. {\bf 6} (1965) 918.

\bibitem{7} D.Y. Xu, Class. Quantum Gravity {\bf 5} (1988) 871.

\bibitem{8} H. Kim, Phys. Rev. {\bf D 59} (1999) 064002.

\bibitem{9} S. Yazadjiev, Gen. Relativ. Gravit. {\bf 32} (2000) 2345.

\bibitem{10} M. Azreg-A\"{i}nou, Eur. Phys. J. {\bf C 74} (2014) 2865.

\bibitem{11} L. Herrera and J. Jimenez, J. Math. Phys.{\bf 23} (1982) 2339.

\bibitem{12} S.P. Drake and R. Turolla, Class. Quantum Gravity {\bf 14} (1997) 1883.

\bibitem{13} N. Ibohal, Gen. Relativ. Gravit. {bf 37} (2005) 19.
 
\bibitem{14} M. Azreg-A\"{i}nou, Phys. Rev. {\bf D 90} (2014) 064041.

\bibitem{GUP1} A. Kempf, G. Mangano and R. B. Mann, Phys. Rev. {\bf D 52} (1995) 1108.

\bibitem{15} S.W. Hawking, Commun. Math. Phys. \textbf{43} (1975) 199.

\bibitem{16} M. Sharif and W. Javed, Eur. Phys. J. {\bf C 72} (2012) 1997.

\bibitem{17} T. Damoar and R. Ruffini, Phys. Rev. \textbf{D 14} (1976) 332.

\bibitem{18} W. Javed, G. Abbas and R. Ali, Eur. Phys. J. {\bf C 77} (2017) 296.

\bibitem{19} A. \"{O}vg\"{u}n, W. Javed and R. Ali, Advance in High Energy Physics. {\bf 2018} (2018) 11.

\bibitem{20}  W. Javed, R. Ali and G. Abbas, Can. J. Phys. {\bf 97} (2018) 176.

\bibitem{21} W. Javed, R. Ali, R. Babar, A. \"{O}vg\"{u}n, Eur. Phys. J. Plus {\bf 134} (2019) 511.

\bibitem{22} W. Javed, R. Ali, R. Babar, A. \"{O}vg\"{u}n, Chinese Physics {\bf C 144} (2020) 015104.

\bibitem{23} G. Johnson, JHEP {\bf 03} (2020) 038.

\bibitem{24} R. Ali, K. Bamba and S.A.A. Shah, Symmetry. {\bf 631} (2019) 11.

\bibitem{25} R. Ali, K. Bamba, M. Asgher, M.F. Malik 
and S.A.A. Shah, Symmetry. {\bf 1165} (2020) 12.

\bibitem{26} R. Ali, M. Asgher and M.F. Malik, Mod. Phys. Lett. {\bf A 35} (2020) 2050225.

\bibitem{27} M. Rizwan, M.Z. Ali and  A. \"{O}vg\"{u}n,  Mod. Phys. Lett. {\bf A 34} (2019) 1950184.

\bibitem{28} R. Ali, K. Bamba, M. Asgher and S.A.A. Shah, Int. J. Mod. Phys. {\bf D 30} (2021) 2150002.

\bibitem{28a} R. Ali, K. Bamba, S.A.A. Shah
and M.J. Saleem,, Int. J. Mod. Phys. {\bf D 31} (2022) 2250069.

\bibitem{29} W. Javed, R. Babar and A. \"{O}vg\"{u}n, Mod. Phys. Lett. {\bf A 34} (2019) 1950057.

\bibitem{30} W. Javed and R. Babar, Chin. J. Phys. \textbf{61} (2019) 138.

\bibitem{31} R. Babar, W. Javed and A. \"{O}vg\"{u}n, Mod. Phys. Lett. {\bf A 35} (2020) 2050104.

\bibitem{32} A. Kempf, J. Phys. \textbf{A 30} (1997) 2093.

\bibitem{33}  F. Brau, J. Phys. \textbf{A 32} (1999) 7691.

\bibitem{34}  A.F. Ali, S. Das and E.C. Vagenas, Phys. Lett. \textbf{B 678} (2009) 497.

\bibitem{35} A. \"{O}vg\"{u}n and K. Jusufi, Eur. Phys. J. Plus {\bf 132} (2017) 298.

\bibitem{z3} M. Faizal and M.M. Khalil, Int. J. Mod. Phys.  \textbf{A 30} (2015) 1550144.

\bibitem{z4} B. Pourhassan and M. Faizal, Nucl. Phys. \textbf{B 913} (2016) 834.

\bibitem{z5} A. Jawad and M.U. Shahzad, Eur. Phys. J. \textbf{C 77} (2017) 349.

\bibitem{z6} M. Zhang, Nucl. Phys. \textbf{B 935} (2018) 170.

\bibitem{z13} K. Ghaderi and B. Malakolkalami, Nucl. Phys. \textbf{B 903} (2016) 10.

\bibitem{z14a} R. Ali, R. Babar, M. Asgher and
S.A.A. Shah, Annals of Physics, {\bf 432} (2021) 168572.

\bibitem{z14b} R. Ali and M. Asgher, New Astronomy {\bf 93} (2022) 101759.

\bibitem{z14c} R. Ali, R. Babar, M. Asgher and S.A.A. Shah, Int. J. Geom. Methods Mod. Phys. {\bf 19} (2022) 2250017.

\bibitem{z15}  M. Sharif and  Z. Akhtar, Phys. Dark Universe {\bf 29} (2020) 100589.

\bibitem{z16} M. Sharif and Z. Akhtar, Chin. J. Phys {\bf 71} (2021) 669.

\bibitem{z17} W. Javed, et al,  Mod. Phys. Lett. {\bf A 33} (2018) 1850089.

\bibitem{z18} Z. Yousaf, et al, Int. J. Geom. Methods Mod. {\bf 19} (2022) 2250102.

\bibitem{RB1} R. Babar, M. Asgher and R. Ali, Phys. Scr. {\bf 97}(2022)125201.

\bibitem{RB2} Z. Akhtar, A. Khan, Z. Ahmad, R. Ali, New Astronomy {\bf 98}(2023)101936.

\bibitem{S1} J. Wess and J. Bagger, Supersymmetry and Supergravity, Princeton University Press, 1992.

\bibitem{S2} G.W. Gibbons, Supersymmetric Soliton States in Extended Supergravity
Theories, in: P. Breitenlohner and H.P. D\"{u}rr (eds.), Unified
Theories of Elementary Particles, Lecture Notes in Physics 160,
Springer, 1982.

\bibitem{S3} A. Gallerati, Int. J. Mod. Phys. A {\bf 34}(2019) 1930017.

\bibitem{36} M. Cvetic and S.S. Gubser, JHEP {\bf 04} (1999) 024.

\bibitem{37} K. Behrndt, M. Cvetic and W.A. Sabra, Nucl. Phys. {\bf B 553} (1999) 317.

\bibitem{R6} Z. Xu, X. Hou, X Gong and J. Wang Eur. Phys. J. {\bf C 78}(2018) 513.

\bibitem{R7} H.C.D.L. Junior, L.C.B. Crispino, P.V.P. Cunha and C.A.R. Herdeiro, Eur. Phys. J. {\bf C 80}(2020) 1036.

\bibitem{R8} C.Y. CHEN, Int. J. Geom. Methods Mod. {\bf 19}(2022) 2250176.

\bibitem{R9} R.C. Pantig and E.T. Rodulfo, Chin. J. Phys.
{\bf 68}(2020) 236.

\bibitem{z7} P. Pradhan, Universe \textbf{5} (2019) 57.

\end{thebibliography}
\end{document}